\documentclass[twocolumn,prd,amsmath,superscriptaddress,nofootinbib,floatfix]{revtex4}

\usepackage{amsmath}
\usepackage{graphics}
\usepackage{color}

\def\gsim{\;\rlap{\lower 2.5pt
 \hbox{$\sim$}}\raise 1.5pt\hbox{$>$}\;}
\def\lsim{\;\rlap{\lower 2.5pt
 \hbox{$\sim$}}\raise 1.5pt\hbox{$<$}\;}

\def\be{\begin{equation}}
\def\ee{\end{equation}}
\def\bea{\begin{eqnarray}}
\def\eea{\end{eqnarray}}

\begin{document}

\title{Challenges to the DGP Model from Horizon-Scale Growth
and Geometry}

\author{Wenjuan Fang}
\affiliation{Department of Physics, Columbia University, New York,
NY 10027}
\author{Sheng Wang}
\affiliation{Kavli Institute for Cosmological Physics, Enrico Fermi
Institute, University of Chicago, Chicago, IL 60637}
\author{Wayne Hu}
\affiliation{Kavli Institute for Cosmological Physics, Enrico Fermi
Institute, University of Chicago, Chicago, IL 60637}
\affiliation{Department of Astronomy and Astrophysics, University of
Chicago, Chicago, IL 60637}
\author{Zolt\'{a}n Haiman}
\affiliation{Department of Astronomy, Columbia University, New York,
NY 10027}
\author{Lam Hui}
\affiliation{Department of Physics, Columbia University, New York,
NY 10027}
\author{Morgan May}
\affiliation{Brookhaven National Laboratory, Upton, NY 11973}

\date{\today}

\begin{abstract}
We conduct a Markov Chain Monte Carlo study of the
Dvali-Gabadadze-Porrati (DGP) self-accelerating braneworld scenario
given the cosmic microwave background (CMB) anisotropy, supernovae and
Hubble constant data by implementing an effective dark energy
prescription for modified gravity into a standard Einstein-Boltzmann
code.  We find no way to alleviate the tension between distance measures
and horizon scale growth in this model.  Growth alterations due to
perturbations propagating into the bulk appear as excess CMB anisotropy
at the lowest multipoles.  In a flat cosmology, the maximum likelihood
DGP model is nominally a $5.3\sigma$ poorer fit than $\Lambda$CDM.
Curvature can reduce the tension between distance measures but only at
the expense of exacerbating the problem with growth leading to a
$4.8\sigma$ result that is dominated by the low multipole CMB
temperature spectrum.  While changing the initial conditions to reduce
large scale power can flatten the temperature spectrum, this also
suppresses the large angle polarization spectrum in violation of recent
results from WMAP5.  The failure of this model highlights the power of
combining growth and distance measures in cosmology as a test of gravity
on the largest scales.
\end{abstract}

\maketitle

\section{Introduction}

On the self-accelerating branch of the Dvali-Gabadadze-Porrati (DGP)
braneworld model \cite{DGP}, cosmic acceleration arises from a
modification to gravity at large scales rather than by introducing a
form of mysterious dark energy with negative pressure \cite{De01}.  In
this model our universe is a ($3+1$)-dimensional brane embedded in an
infinite Minkowski bulk with differing effective strengths of gravity.
The relative strengths define a crossover scale on the brane beyond
which ($4+1$)-dimensional gravity and bulk phenomena become important.

A number of theoretical and observational problems of the
self-accelerating branch of DGP have been recently uncovered. This
branch suffers from pathologies related to the appearance of ghost
degrees of freedom
\cite{LuPR03,NicR04,Ko04,GoKS06,ChGNP06,DefGI06,Dv06,KoyS07}. Ghosts
can lead to runaway excitations when coupled to normal modes and
their existence can invalidate the self-accelerating background
solution as well as linear perturbations around it.

Observational problems also arise if one posits that ghosts and strong
coupling do not invalidate the idea of self-acceleration itself, {\it
i.e.}~that our Hubble volume behaves in a manner that is perturbatively
close on scales near the horizon to the modified Friedmann equation
specified on this branch.  These problems fall into two classes, those
due to the background expansion history and those due to the growth of
structure during the acceleration epoch.

In a flat spatial geometry the DGP model adds only one degree of freedom
to fit acceleration, the crossover scale.  Like $\Lambda$CDM, the extra
degree of freedom can be phrased as the matter density relative to the
critical density.  Remarkably, with this one parameter, the $\Lambda$CDM
model can fit three disparate sets of distance measures: the local
Hubble constant and baryon acoustic oscillations, the relative distances
to high-redshift supernovae (SNe), and the acoustic peaks in the cosmic
microwave background (CMB).  The flat DGP model cannot fit these
observations simultaneously and the significance of the discrepancy
continues to grow.  The addition of spatial curvature can help alleviate
some, but not all, of this tension \cite{FaG06,MaM06,SoSH07}.

On intermediate scales that encompass measurements of the large scale
structure of the Universe, the early onset of modifications to the
expansion also imply a substantial reduction in the linear growth of
structure which is itself slightly reduced by the modification to
gravity \cite{LuSS04,KoM06}.  If extended to the mildly nonlinear regime
where strong coupling effects may alter growth, this reduction is in
substantial conflict with weak lensing data \cite{WaHMH07}.

On scales approaching the crossover scale, which are probed by the
CMB, the unique modification to gravity in the DGP model itself
strongly alters the growth rate.  Here the propagation of
perturbations into the bulk requires a ($4+1$)-dimensional
perturbation framework \cite{De02}.  The approximate iterative
solutions introduced in Rfn.~\cite{SaSH07} have been recently
verified to be sufficiently accurate by a more direct calculation
\cite{CaKSS08} but still are computationally too expensive for Monte
Carlo explorations of the DGP parameter space.  More recently, a
($3+1$)-dimensional effective approach dubbed the parameterized
post-Friedmann (PPF) framework \cite{Hu08,HuS07} has been developed
that accurately encapsulates modified gravity effects with a closed
effective dark energy system.  The PPF approach enables standard
cosmological tools such as an Einstein-Boltzmann linear theory
solver to be applied to the DGP model.

In this {\it Paper}, we implement the PPF approach to DGP and
conduct a thorough study of the tension between CMB distance, energy
density, and growth measures, along with the SNe and local distance
measures, across an extended DGP parameter space.  We find that even
adding epicycles to the DGP model does not significantly improve the
agreement with the data. Curvature, while able to alleviate the
problem with distances, exacerbates the problem with growth.
Changing the initial power spectrum to remove excess power in the
temperature spectrum destroys the agreement with recent polarization
measurements from the five-year Wilkinson Microwave Anisotropy Probe
(WMAP) \cite{Noetal08}.

The outline of the paper is as follows.  In \S \ref{sec:theory} we
review the impact of the DGP modifications on distance measures and
the growth of structure emphasizing an effective dark energy PPF
approach that is detailed in Appendix A and compared with a
growth-geometry splitting approach in Appendix B.  We present the
results of the likelihood analysis in \S \ref{sec:results} and
discuss these results in \S \ref{sec:discuss}.

\section{DGP Distance and Growth Phenomenology}
\label{sec:theory}

In the DGP model, gravity remains a metric theory on the brane in a
background that is statistically homogeneous and isotropic.  Its
modifications therefore are confined to the field equations that relate
the metric to the matter.  Once the metric is obtained, all the usual
implications for the propagation of light from distant sources and the
motion of matter remain unchanged.  In this section, we review the DGP
modifications to the background metric, or expansion history, and the
gravitational potentials, or linear metric perturbations.  We cast these
modifications in the language of an effective dark energy contribution
under ordinary gravity \cite{Hu08,HuS07,Ba07,KuS07,CaCM07,JaZ07,BeZ08}.

\subsection{Background Evolution}
\label{subsec:background}

That the DGP model is a metric theory in a statistically homogeneous
and isotropic universe imposes a background
Friedmann-Robertson-Walker (FRW) metric on the brane.  The FRW
metric is specified by two quantities, the evolution of the scale
factor $a$ and the spatial curvature $K$.  The DGP model modifies
the field equation, {\it i.e.}~the Friedmann equation, relating the
evolution of the scale factor $H=a^{-1} da/dt$ to the matter-energy
content.  On the self-accelerating branch it becomes (see, {\it
e.g.}, Rfns.~\cite{De01,DeLRZA02})
\begin{equation}
H^2  = \left(\sqrt{\frac{8 \pi G}{3}\sum_i
\rho_i+\frac{1}{4r_c^2}}+\frac{1}{2r_c}\right)^2-\frac{K}{a^2},
\label{eqn:modfriedmann}
\end{equation}
where $r_c$ is the crossover scale and subscript $i$ labels the true
matter-energy components of the universe to be distinguished below from
the effective dark energy contribution.

On the other hand, given the metric the matter evolves in the same way
as in ordinary gravity
\begin{equation}
\dot{\rho}_i=-3aH(\rho_i+P_i).
\label{eqn:densityevolution}
\end{equation}
Overdots represent derivatives with respect to conformal time $\eta
= \int dt/a$.  From these relations, one sees that as $a \rightarrow
\infty$, $H \rightarrow r_c^{-1}$ and the Universe enters a de Sitter
phase of accelerated expansion.

In the limit that $r_c \rightarrow \infty$ the ordinary Friedmann
equation is recovered.  The effect of a finite $r_c$ compared with the
Hubble scale can be encapsulated in a dimensionless parameter
$\Omega_{r_c}$ much like the usual contributions of the density
$\Omega_i = 8\pi G\rho_i/3H_0^2$ and curvature $\Omega_K = -K/H_0^2$ to
the expansion rate
\begin{equation}
\Omega_{r_c} \equiv \frac{1}{4r_c^2H_0^2} .
\end{equation}
The modified Friedmann equation (\ref{eqn:modfriedmann}) today then
becomes the constraint equation
\begin{equation}
1=\left(\sqrt{\Omega_{r_c}} +\sqrt{\Omega_{r_c}+\sum_i
\Omega_i}~\right)^2+\Omega_K.
\label{eqn:normDGP}
\end{equation}

It is convenient and instructive to recast the impact of $r_c$ as an
effective dark energy component.  With the same background evolution,
the effective dark energy will have an energy density of
\begin{equation}
\rho_e \equiv \frac{3}{8 \pi G} \left(H^2 +\frac{K}{a^2}\right)-
\sum_i \rho_i.
\end{equation}
Conservation of its energy-momentum tensor is guaranteed by the
Bianchi identities and requires its ``equation of state'' $w_e
\equiv P_e/\rho_e$ to be given by Eq.~(\ref{eqn:densityevolution})
\begin{equation}
w_e=\frac{\sum_i(\rho_i+P_i)}{3(H^2 + a^{-2}K)/8\pi G + \sum_i
\rho_i}-1.
\label{eqn:we}
\end{equation}
If we define $\Omega_e$ in the same way as $\Omega_i$, the usual
constraint condition applies $\sum_i\Omega_i +\Omega_e + \Omega_K=1$.
Comparing it to Eq.~(\ref{eqn:normDGP}), we obtain the following
relationship between $\Omega_e$ and $\Omega_{r_c}$
\begin{equation}
\Omega_e=2\sqrt{\Omega_{r_c}(1-\Omega_K)}.
\label{eqn:omegae}
\end{equation}

Given $w_e$ and $\Omega_e$, we can now describe the background evolution
of the DGP cosmology by using the ordinary Friedmann equation for $H$.
Likewise the Hubble parameter specifies the comoving radial distance
\begin{equation}
D(z)=\int_0^z\frac{dz'}{H(z')}\,,
\end{equation}
and the luminosity distance
\begin{equation}
d_L(z)=\frac{(1+z)}{\sqrt{-\Omega_K}H_0}\sin{\left(\sqrt{-\Omega_K}H_0\,
D(z)\right)}\,,
\label{eqn:luminositydistance}
\end{equation}
as usual.

\begin{figure}[t!]
\resizebox{90mm}{!}{\includegraphics{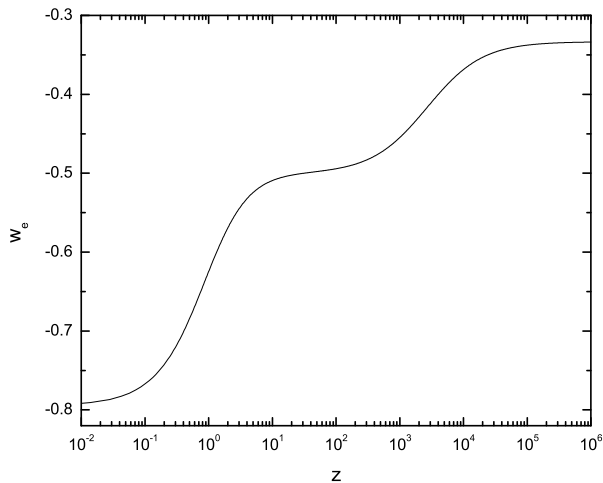}}
\caption{Equation of state of the effective dark energy $w_e$ for the
self-accelerating DGP model with $\Omega_m=0.26$ and $\Omega_K=0$.}
\label{fig:we}
\end{figure}

The effective dark energy for DGP has quite a different equation of
state from that of a cosmological constant. For example, when there
is no curvature, $w_e$ starts at ${-1/(1+\Omega_m)}$ at the present,
approaches $-1/2$ in the matter dominated regime and $-1/3$ in the
radiation dominated regime (see Fig.~\ref{fig:we}).

With the same values of parameters, DGP will have a larger amount of
effective dark energy at a given redshift as $w_e$ is always larger than
$-1$, and the universe will expand at a larger rate than in
$\Lambda$CDM.  This reduces the absolute comoving radial distance to a
distant source.  For relative distance measures like the SNe, the
reduction in $H_0 d_L(z)$ can be compensated by lowering the {\it
fractional} contribution of matter through $\Omega_m$.  The same is not
true for absolute distances, such as those measured by the CMB, baryon
oscillations, and Hubble constant, if the overall matter contribution to
the expansion rate $\Omega_m H_0^2 a^{-3}$ remains fixed.

\subsection{Structure Formation in the Linear Regime}

The same methodology of introducing an effective dark energy component
for the background expansion history applies as well to the linear
metric perturbations that govern the evolution of large scale structure.

To define the effective dark energy, one must first parameterize the
solutions to the ($4+1$)-dimensional equations involving metric
perturbations in the brane as well as the bulk \cite{De02}.  Three
simplifications aid in this parameterization \cite{HuS07}.  The first is
that at high redshift the effect of $r_c$ and the extra dimension goes
away rapidly [see Eq.~(\ref{eqn:modfriedmann})].  The parameterization
needs only to be accurate between the matter dominated regime and the
present.

Likewise, the predictions of $\Lambda$CDM for the relationships
between the matter and baryon density at recombination and
morphology of the CMB acoustic peaks remain unchanged.  As a
consequence, the shape of the CMB acoustic peaks still constrains
the physical cold dark matter and baryon density $\Omega_c h^2$ and
$\Omega_b h^2$ as usual.  With these as fundamental high redshift
parameters, in a flat spatial geometry, the DGP degree of freedom
$r_c$ is then specified by either
$\Omega_m=1-\Omega_e=\Omega_c+\Omega_b$ or $H_0$.  We shall see
below that in a flat universe the competing requirements of CMB and
SNe distance measures on $r_c$ will slightly shift the values of
$\Omega_c h^2$ and $\Omega_b h^2$ to reach a compromise at the
expense of the goodness of fit.

The second simplification is that well below the horizon, perturbations
on the brane are in the quasi-static regime where time derivatives can
be neglected in comparison with spatial gradients and propagation
effects into the bulk are negligible.  This allows the equations to be
simply closed on the brane by a modified Poisson equation (\ref{eqn:qs})
that can be recast as arising from the anisotropic stress of the
effective dark energy \cite{LuSS04,KoM06}.

The final simplification is that on scales above the horizon, the impact
of the bulk perturbations on the brane becomes scale free and depends
only on time \cite{SaSH07} through dimensionless combinations of $H$,
$r_c$ and $K$.  Furthermore on these scales, generic modifications to
gravity are fully defined by the Friedmann equation and the anisotropic
stress of the effective dark energy \cite{HuE98,Be06}.

As shown in Rfn.~\cite{HuS07} and detailed in Appendix A,
interpolation between these two limits leads to a simple PPF
parameterization of DGP on all linear scales.  This parameterization
has been verified to be accurate at a level substantially better
than required by cosmic variance by a direct computation of bulk
perturbations \cite{CaKSS08}.  With such a parameterization,
efficient Einstein-Boltzmann codes such as CAMB \cite{camb} can be
modified to calculate the full range of CMB anisotropy, as is done
in this paper.

The result of this calculation is that compared with $\Lambda$CDM,
the growth of structure in the DGP model during the acceleration
epoch is suppressed.  In the quasi-static regime, this suppression
is mostly due to the higher redshift extent of the acceleration
epoch discussed in the previous section (see also Fig.~\ref{fig:we})
with a small component from the effective anisotropic stress or
modification to the Poisson equation \cite{LuSS04,KoM06}.  On scales
approaching $r_c$, the effect of the anisotropic stress becomes much
more substantial due to the perturbations propagating into the bulk
\cite{SaSH07} resulting in an even stronger integrated Sachs-Wolfe
(ISW) effect in the CMB anisotropy power at the lowest multipoles
\cite{SoSH07}.

\section{Constraints from Current Observations}
\label{sec:results}

In this section, we employ Markov Chain Monte Carlo (MCMC) techniques to
explore constraints on the DGP parameter space from current
observations, and compare them with the successful $\Lambda$CDM model.
The data we use are: the Supernovae Legacy Survey (SNLS) \cite{SNLS},
the CMB anisotropy data from the five-year WMAP \cite{WMAP5} for both
temperature and polarization (TT + EE + TE), and the Hubble constant
measurement from the Hubble Space Telescope (HST) Key Project \cite{H0}.

We use the public MCMC package \textit{CosmoMC} \cite{cosmomc}, with a
modified version of CAMB for DGP described in Appendix A, to sample
the posterior probability distributions of model parameters.  The MCMC
technique employs the Metropolis-Hastings algorithm \cite{Metropolis53,
Hastings70} for the sampling, and the Gelman and Rubin $R$ statistic
\cite{GeR92} for the convergence test. We conservatively require
$R-1<0.01$ for the eight chains we run for each model, and this
generally gave us $\sim 5000$ independent samples.

The SNe magnitude-redshift relation probes the relative luminosity
distance between the low and high redshift sample.  The luminosity
distance itself is completely determined by the background expansion
history through Eq.~(\ref{eqn:luminositydistance}).  On the other
hand the power spectra of the CMB anisotropy probes not only the
background expansion, but also the growth of structure.  To better
separate the two types of information, we also consider a canonical
scalar field (``quintessence'' or QCDM) model with the same
expansion history as DGP. This model is defined by the equation of
state parameter $w_e$ and density parameter $\Omega_e$ in
Eqs.~(\ref{eqn:we}) and (\ref{eqn:omegae}).

With only scalar perturbations in consideration, our basic parameter
set is chosen to be $\{\Omega_b h^2, \Omega_c h^2, \theta_s, \tau,
n_s, A_s\}$, which in turn stand for the density parameters of
baryons and cold dark matter, angular size of the sound horizon at
recombination, optical depth from reionization
(assumed to be instantaneous), spectra index of the primordial
curvature fluctuation and its amplitude at $k_*=0.002$ Mpc$^{-1}$,
{\it i.e.}, $\Delta_{\zeta}^2=A_s(k/k_*)^{n_s-1}$.  Also, we follow
Rfn.~\cite{WMAP5}, and include $A_{SZ}$, with flat prior of
$0<A_{SZ}<2$, to account for the contributions to the CMB power
spectra from Sunyaev-Zeldovich fluctuations. The lensing effect on
the CMB is neglected. Note that in all the three models we
considered, {\it i.e.}~self-accelerating DGP, $\Lambda$CDM and QCDM,
we have the same parameter sets and priors except that we restrict
the DGP parameter space to $H_0 r_c>1.08$ so that metric
fluctuations remain well behaved [see Eq.~(\ref{eqn:gsh})]. We
apply this prior to the QCDM model as well for a fair comparison. In
practice, the excluded models are strongly disfavored by the data
and the prior is only necessary for numerical reasons.

%
\subsection{Flat Models}

We start with the minimal parameterization of a flat universe with
scale free initial conditions.  In this case the three model classes
$\Lambda$CDM, DGP and QCDM all have only one parameter that
describes acceleration.  In the chain parameters this is $\theta_s$
but can be equivalently defined as the derived parameters $H_0$ or
$\Omega_m$. The constraints on the three model classes are given in
Table \ref{tab:fmean} for the means and marginalized errors on
various parameters. Table \ref{tab:fbf} shows the best-fit values of
the parameters and the corresponding likelihoods, which serve as a
``goodness of fit'' criterion.

\begin{table}
\caption{\label{tab:fmean}Mean and marginalized errors for various
parameters of the self-accelerating DGP, QCDM with the same expansion
history as the DGP and $\Lambda$CDM models from SNLS + WMAP5 + HST, {\it
assuming a flat universe}. The first 6 parameters are directly varied
when running the Markov Chains, while the others are derived parameters,
as are in the following tables.}
\begin{tabular}{|c |c |c |c |}
\hline
parameters &  DGP  &  QCDM &  $\Lambda$CDM \\
\hline
$100\Omega_b h^2$    &2.36$\pm$0.07    &2.32$\pm$0.07    &2.25$\pm$0.06   \\
$\Omega_c h^2$       &0.090$\pm$0.005  &0.090$\pm$0.005  &0.109$\pm$0.005 \\
$100\theta_s$        &1.042$\pm$0.003  &1.041$\pm$0.003  &1.040$\pm$0.003 \\
$\tau$               &0.10$\pm$0.02    &0.10$\pm$0.02    &0.09$\pm$0.02   \\
$n_s$                &1.00$\pm$0.02    &0.99$\pm$0.02    &0.96$\pm$0.01   \\
$\ln{[10^{10}A_s]}$  &3.02$\pm$0.05    &3.05$\pm$0.04    &3.18$\pm$0.04   \\
\hline
$H_0$                &$66\pm2$         &65$\pm$2         &72$\pm$2        \\
$\Omega_m$           &$0.26\pm0.02$    &0.26$\pm$0.02    &0.26$\pm$0.02   \\
$\Omega_{r_c}$       &$0.136\pm0.009$  &..               &..              \\
\hline
\end{tabular}
\end{table}
\begin{table}
\caption{\label{tab:fbf}Parameters and the likelihood values at the
best-fit point of the self-accelerating DGP, QCDM with the same
expansion history as the DGP and $\Lambda$CDM models fitting to SNLS +
WMAP5 + HST, {\it assuming a flat universe}.}
\begin{tabular}{|c |c |c |c |}
\hline
parameters &  DGP  &  QCDM &  $\Lambda$CDM \\
\hline
$100\Omega_b h^2$     &2.37    &2.32    &2.26   \\
$\Omega_c h^2$        &0.0888  &0.0907  &0.110  \\
$100\theta_s$         &1.04    &1.04    &1.04   \\
$\tau$                &0.0954  &0.0998  &0.0825 \\
$n_s$                 &0.998   &0.983   &0.959  \\
$\ln {[10^{10}A_s]}$  &3.01    &3.06    &3.18   \\
\hline
$H_0$                 &66.0    &65.1    &71.6   \\
$\Omega_m$            &0.258   &0.269   &0.258  \\
$\Omega_{r_c}$        &0.138   &..      &..     \\
\hline
$-2\ln L$             &2805.8  &2797.6  &2777.8 \\
\hline
\end{tabular}
\end{table}

First, we compare the constraints on the QCDM model with those on
$\Lambda$CDM. The differences are expected to reflect those between
the background expansion histories of the DGP and $\Lambda$CDM
models. In spite of the clustering effects on the largest scales of
a quintessence dark energy, the difference between QCDM and
$\Lambda$CDM is completely encoded in the equation of state of their
dark energy components given the fixed sound speed of quintessence.

Constraints from the SNe magnitudes come from the dimensionless
luminosity distance $H_0 d_L(z)$ [see
Eq.~(\ref{eqn:luminositydistance})], once the unknown absolute
magnitude is marginalized.  In order to match the predictions for
$H_0 d_L(z)$ of a flat $\Lambda$CDM model, we would expect that the
QCDM model has a smaller $\Omega_m$ to compensate the larger $w_e$
its dark energy has.  However, lowering $\Omega_m$ also shortens
more of the distance to the last scattering surface and hence
increases the angular size of the sound horizon (see \S
\ref{subsec:background}).

The physical scale of the sound horizon can partially compensate and
is controlled by $\Omega_c h^2$ and $\Omega_b h^2$ but these
parameters are also well measured by the shape of the peaks and can
only be slightly adjusted at a cost to the goodness of fit.  Thus
the parameter ranges in Table \ref{tab:fmean} for $\Omega_c h^2$
decrease and $\Omega_b h^2$ increase slightly which both have the
effect of decreasing the angular size of the horizon while
$\Omega_m$ remains nearly unchanged.  This compromise between the
energy density and distance constraints results in tension between
the CMB and SNe data.  This tension shows up as a difference between
the $-2\ln L$ values of these two models: $2\ln L(\Lambda{\rm
CDM})-2\ln L({\rm QCDM})\simeq 20$.

\begin{figure}[t!]
\resizebox{90mm}{!}{\includegraphics{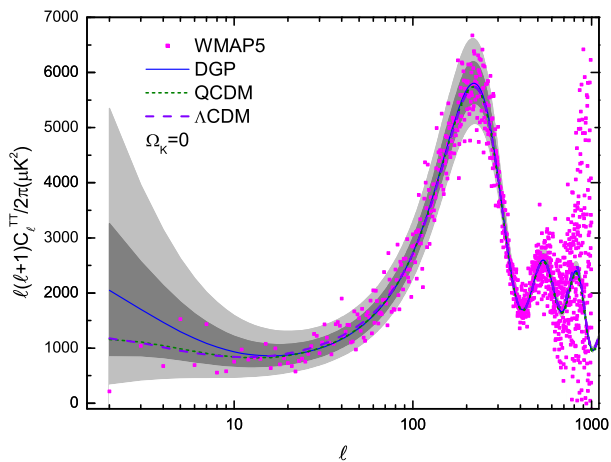}}
\caption{\label{fig:fcmpTT} Predictions for the power spectra of the
CMB temperature anisotropies $C_{\ell}^{\rm TT}$ of the best-fit DGP
(solid), QCDM with the same expansion history as DGP (short-dashed),
and $\Lambda$CDM (dashed, coincident with QCDM at low $\ell$) models
obtained by fitting to SNLS + WMAP5 (both temperature and
polarization) + HST, {\it assuming a flat universe}. Bands represent
the 68\% and 95\% cosmic variance regions for the DGP model.  Points
represent WMAP5 measurements; note that noise dominates over cosmic
variance for $\ell \gtrsim 500$.}
\end{figure}

The QCDM model also favors a larger $n_s$ and a smaller $A_s$.  This
is a consequence of the larger ISW effect in the QCDM model due to a
slower growth rate.  Tilting the spectrum can compensate for the
excess power in the low-$\ell$ modes (as shown by the
near-coincidence of the short-dashed and dashed curves in
Fig.~\ref{fig:fcmpTT}).

Next we compare the constraints on the DGP model with those on QCDM.
Since the two models have the same expansion history, the differences
are entirely caused by the differing growth rates.  Due to the
propagation of perturbations into the bulk for scales near $r_c$ and the
opposite effect of dark energy clustering in QCDM, there is a
substantially stronger ISW effect in the first few multipoles of the CMB
anisotropy power \cite{SoSH07}.  Since this effect is only important on
the largest angular modes in the CMB TT power spectra which is further
limited by the large cosmic variance, the parameter ranges for these two
models do not differ significantly.  Nonetheless from
Fig.~\ref{fig:fcmpTT}, it is clear that the best-fit DGP model over
predicts the low-$\ell$ modes anisotropy, though as before, $n_s$ and
$A_s$ adjustments try to reduce the primordial perturbations on large
scales.  This leads to DGP being an even worse fit than QCDM with
$2\ln L({\rm QCDM})-2\ln L({\rm DGP})\simeq 8$.

When DGP is compared to $\Lambda$CDM, this brings the change in $-2\ln
L$ for the maximum likelihood parameters to $\simeq 28$, where $\sim
70\%$ is driven by the background expansion, while $\sim 30\%$ by the
dynamical effects on structure growth.

\subsection{Adding in Curvature}
\label{subsec:omk}

From our analysis in the above section, the flat DGP model is a poor
fit to the current observations mostly because it cannot
simultaneously satisfy the geometrical requirements of the relative
luminosity distances of the SNe and the angular size of the sound
horizon at recombination with a single parameter.  Since curvature
has more of an effect on high redshift distance measures, the
tension in the distance measures can be alleviated by including
$\Omega_K$ in the parameter space \cite{MaM06,SoSH07}.  Our results
are given in Table \ref{tab:nfmean} and Table \ref{tab:nfbf}.

\begin{table}
\caption{\label{tab:nfmean} Mean and marginalized errors for various
parameters of the self-accelerating DGP, QCDM with the same expansion
history as the DGP and $\Lambda$CDM models from SNLS + WMAP5 + HST, {\it
allowing curvature}.}
\begin{tabular}{|c |c |c |c |}
\hline
parameters &  DGP  &  QCDM &  $\Lambda$CDM \\
\hline
$100\Omega_b h^2$     &2.37$\pm$0.07    &2.34$\pm$0.07    &2.25$\pm$0.06    \\
$\Omega_c h^2$        &0.096$\pm$0.006  &0.098$\pm$0.006  &0.108$\pm$0.006  \\
$100\theta_s$         &1.043$\pm$0.003  &1.042$\pm$0.003  &1.040$\pm$0.003  \\
$\tau$                &0.09$\pm$0.02    &0.09$\pm$0.02    &0.09$\pm$0.02    \\
$\Omega_K$            &0.019$\pm$0.008  &0.027$\pm$0.008  &-0.004$\pm$0.009 \\
$n_s$                 &1.00$\pm$0.02    &0.99$\pm$0.02    &0.96$\pm$0.01    \\
$\ln {[10^{10}A_s]}$  &3.02$\pm$0.05    &3.06$\pm$0.04    &3.18$\pm$0.04    \\
\hline
$H_0$                 &74$\pm$4         &77$\pm$5         &70$\pm$4         \\
$\Omega_m$            &0.22$\pm$0.03    &0.20$\pm$0.03    &0.27$\pm$0.03    \\
$\Omega_{r_c}$        &0.15$\pm$0.01    &..               &..               \\
\hline
\end{tabular}
\end{table}
\begin{table}
\caption{\label{tab:nfbf} Parameters and likelihood values at the
best-fit point of the self-accelerating DGP, QCDM with the same
expansion history as the DGP and $\Lambda$CDM models fitting to SNLS +
WMAP5 + HST, {\it allowing curvature}.}
\begin{tabular}{|c |c |c |c |}
\hline
parameters &  DGP  &  QCDM &  $\Lambda$CDM \\
\hline
$100\Omega_b h^2$     &2.38    &2.36    &2.27     \\
$\Omega_c h^2$        &0.0937  &0.0960  &0.107    \\
$100\theta_s$         &1.04    &1.04    &1.04     \\
$\tau$                &0.0887  &0.0914  &0.0884   \\
$\Omega_K$            &0.0189  &0.0268  &-0.00553 \\
$n_s$                 &0.996   &0.992   &0.959    \\
$\ln {[10^{10}A_s]}$  &3.02    &3.05    &3.18     \\
\hline
$H_0$                 &73.8    &78.3    &69.8     \\
$\Omega_m$            &0.216   &0.195   &0.266    \\
$\Omega_{r_c}$        &0.149   &..      &..       \\
\hline
$-2\ln L$             &2800.8  &2787.2  &2777.5   \\
\hline
\end{tabular}
\end{table}

With curvature, the $-2\ln L$ of the maximum likelihood $\Lambda$CDM
model almost has no improvement. This is consistent with the results
of Rfn.~\cite{Ko08}, who found strong limits on curvature in
$\Lambda$CDM, by fitting WMAP5 data combined with SNe or HST.  As
expected the maximum likelihood model in the QCDM space improves in
$-2\ln L$ by $\sim10$. The QCDM model needs an open universe to
increase the distance to last scattering to compensate the smaller
$\Omega_m$, consistent with the findings by \cite{MaM06}.  Even with
this additional freedom, distance measures remain in slight tension
due to the Hubble constant since lowering $\Omega_m$ with $\Omega_m
h^2$ well determined by the CMB implies a higher Hubble constant
\cite{SoSH07}.  The allowed amount of this shift is limited by the
HST Key Project constraint of $H_0=72 \pm 8$ km s$^{-1}$ Mpc$^{-1}$.
For QCDM the baryon acoustic oscillation constraint would already
disfavor such a shift \cite{detection,hutsib,percival,gch08} but its
application to DGP requires cosmological simulations of the strong
coupling regime. Note that with the distance tension partially
removed, the shifts in $\Omega_c h^2$ and $\Omega_b h^2$ are
reduced.

\begin{figure}[t!]
\resizebox{90mm}{!}{\includegraphics{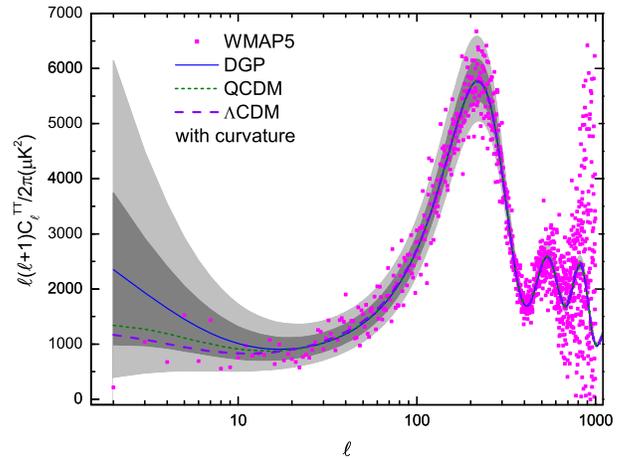}}
\caption{\label{fig:nfcmpTT} Predictions for the power spectra of the
CMB temperature anisotropies $C_{\ell}^{\rm TT}$ of the best-fit DGP
(solid), QCDM with the same expansion history as DGP (short-dashed), and
$\Lambda$CDM (dashed) models obtained by fitting to SNLS + WMAP5 (both
temperature and polarization) + HST, {\it allowing curvature}.}
\end{figure}

The lowering of $\Omega_m$, in addition to the large equation of state
parameter of the quintessence, also causes matter domination to
terminate at an earlier redshift, and leads to a stronger ISW effect in
the QCDM model, as can be seen in Fig.~\ref{fig:nfcmpTT}, again with a
partial compensation from $n_s$ and $A_s$.  The net difference of $-2\ln
L$ values of the QCDM model compared to $\Lambda$CDM is $\simeq 10$,
50\% smaller than before but still a significantly poorer fit.

The situation is even worse for DGP.  Here the enhancement of the ISW
effect at low $\Omega_m$ is even more substantial.  Thus the mean value
of $\Omega_K$ is smaller than the optimal one for the distance
constraints in QCDM.  Even adjusting the other parameters to give the
maximum likelihood model shown in Fig.~\ref{fig:nfcmpTT}, the poor fit
is noticeable at the low multipoles.  For example the probability of
obtaining a quadrupole as extreme as the observations from the DGP
maximum likelihood model is $\sim 1\%$ compared with $\sim 6\%$ for the
$\Lambda$CDM maximum likelihood model.  The net difference in $-2\ln L$
by including curvature as a parameter is only $\sim 5$ in DGP showing
only marginal evidence for curvature in the model at best.

Moreover, the difference from $\Lambda$CDM remains substantial with
$-2\Delta\ln L \simeq 23$.  Since the difference of $-2\Delta\ln L$ from
QCDM can be attributed to the ISW effect, $\sim 40\%$ of this difference
is driven by the background expansion, and $\sim 60\%$ by the
dynamical effects on structure growth.

\subsection{Changing the Initial Power}

The adjustments of $A_s$ and $n_s$ in the above examples suggest that
perhaps a more radical change in the initial power spectrum can bring
DGP back in agreement with the data.  For example, one can sharply
reduce large scale power in the temperature spectrum by cutting off the
initial power spectrum on large scales, below a wavenumber $k_{\rm
min}$.  While this is a radical modification and has no particular
physical motivation, it is useful to check whether such a loss of power
could satisfy the joint temperature and polarization constraints.

CMB polarization at these scales arises from reionization which
occurs at a substantially higher redshift than DGP modifications
affect.  A finite polarization requires not only ionization but also
large scale anisotropy at this epoch.  Eliminating initial power on
these scales eliminates the polarization as well.  The EE power in
the low multipoles has now been measured at 4-$5\sigma$ level
\cite{Noetal08} leading to a significant discrepancy if the large
scale power is removed in the model.

Given that including this parameter does not improve the fit, instead of
adding it to the MCMC parameter space, we illustrate its effects on the
maximum likelihood DGP model in \S \ref{subsec:omk}.

\begin{figure}[t!]
\resizebox{90mm}{!}{\includegraphics{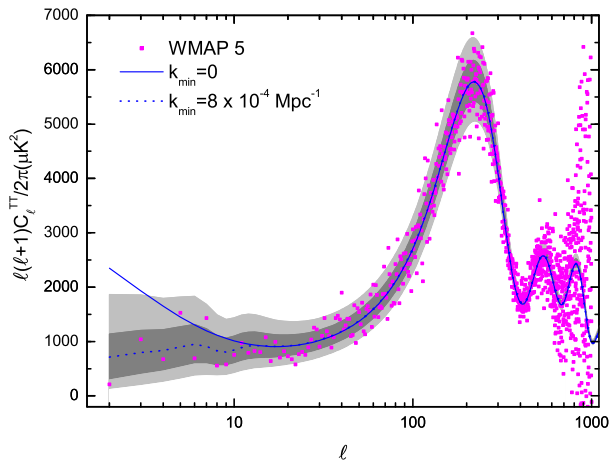}}
\caption{\label{fig:TTtrunc} Predictions for the power spectra of the
CMB temperature anisotropies $C_{\ell}^{\rm TT}$ of the best-fit DGP
model as found in \S \ref{subsec:omk} without cutting off any large
scale primordial perturbations (solid) and with a cut-off scale of
$k_{\rm min}=8\times 10^{-4}$ Mpc$^{-1}$ (dotted) -- the best-fit scale
obtained when fitting to the WMAP 5 year TT data alone, while all other
parameters are fixed at their best-fit values with $k_{\rm min}=0$.}
\end{figure}

By maximizing the likelihood for this model to the TT power alone,
we find the best agreement is obtained at $k_{\rm min}=8\times
10^{-4}$ Mpc$^{-1}$, with $-2\Delta \ln L^{\rm TT} \simeq -12$
compared to $k_{\rm min}=0$. Model predictions with this $k_{\rm
min}$ are plotted in Fig.~\ref{fig:TTtrunc}, together with the WMAP
5 year data.

\begin{figure}[t!]
\resizebox{90mm}{!}{\includegraphics{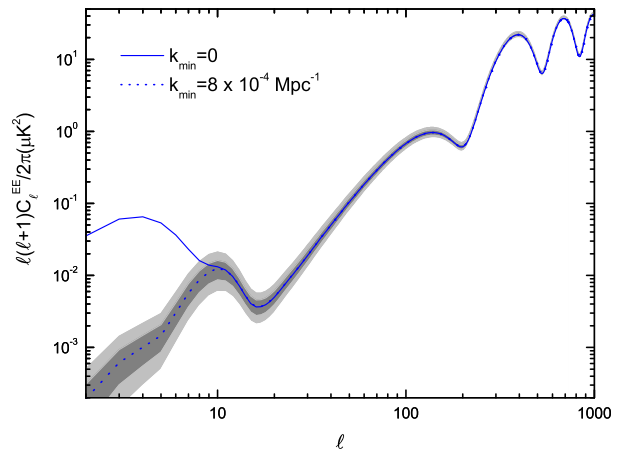}}
\caption{\label{fig:EEtrunc} Predictions for the power spectra of the
CMB E-mode polarization $C_{\ell}^{\rm EE}$ of the best-fit DGP model as
found in \S \ref{subsec:omk} without cutting off any large scale
primordial perturbations (solid) and with a cut-off scale of $k_{\rm
min}=8\times 10^{-4}$ Mpc$^{-1}$ (dotted) -- the best-fit scale obtained
when fitting to the WMAP 5 year TT data alone, while all other
parameters are fixed their best-fit values.  Note here, according to
Rfn.~\cite{Noetal08}, the reionization feature at the lowest-$\ell$
modes is preferred by the data through $\Delta \chi^2=19.6$.}
\end{figure}

\begin{figure}[t!]
\resizebox{90mm}{!}{\includegraphics{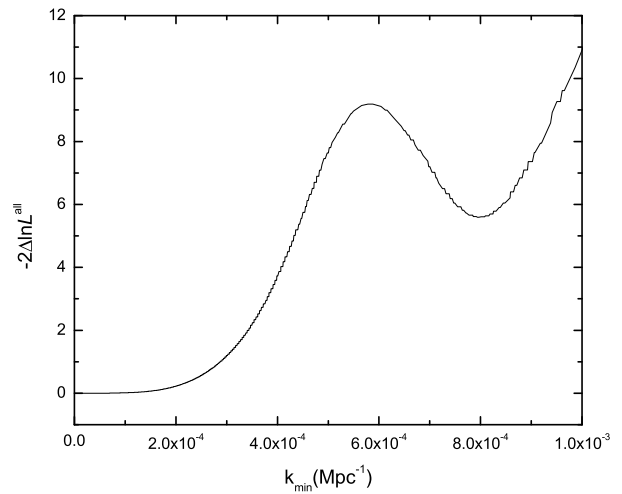}}
\caption{\label{fig:Ltrunc} The total log-likelihood of TT + TE + EE
as a function of the cut-off scale $k_{\rm min}$ of the primordial
perturbations, shown as the difference from its value at $k_{\rm
min}=0$.  We find that when polarization is included, the combined
data does not favor any positive values of $k_{\rm min}$. The local
minimum shows up at the scale of $k_{\rm min}=8\times 10^{-4}$
Mpc$^{-1}$, which is the one favored by the TT data alone. Note all
other parameters are fixed at their best-fit values of the DGP model
as found in \S \ref{subsec:omk}.}
\end{figure}

With this power truncation the over prediction problem for the
low-$\ell$ TT power is alleviated, but the EE polarization power on
large angular scales is significantly reduced (see
Fig.~\ref{fig:EEtrunc}).  When the polarization data of TE + EE are
included, we find $-2\Delta \ln L^{\rm all} \simeq 6$ at $k_{\rm
min}=8\times 10^{-4}$ Mpc$^{-1}$, and when $k_{\rm min}$ is varied,
the combined data does not favor any positive values of $k_{\rm
min}$ as shown in Fig.~~\ref{fig:Ltrunc}.  We conclude that though
the DGP model can be made a better fit to the TT power spectrum with
a large scale cut-off, polarization measurements are now
sufficiently strong to rule out this possibility.  While we have
only included instantaneous reionization models, changing the
ionization history to have an extended high redshift tail can only
exacerbate this problem by increasing EE power in the $\ell \sim
10-30$ regime \cite{MorH08}.

\section{Discussion}
\label{sec:discuss}

We have conducted a thorough Markov Chain Monte Carlo likelihood study
of the parameter space available to the DGP self-accelerating braneworld
scenario given CMB, SNe and Hubble constant data.  To carry out
this study, we have introduced techniques for characterizing modified
gravity and non-canonical dark energy candidates with the public
Einstein-Boltzmann code CAMB that are of interest beyond the DGP
calculations themselves.

We find no way to alleviate substantially the tension between
distance measures and the growth of horizon scale fluctuations that
impact the low multipole CMB temperature and its relationship to the
polarization.  In particular we show that the maximum likelihood
flat DGP model is a poorer fit than $\Lambda$CDM by $2\Delta \ln L =
28$, nominally a $(2\Delta\ln L)^{1/2}=5.3\sigma$ result.
Interestingly, a substantial ($\sim 30\%$) contribution comes from
the change in the growth near the crossover scale where
perturbations leak into the bulk.

Adding in spatial curvature to the model can bring the distance measures
back in agreement with the data but only at the cost of exacerbating
the problem with growth.  The net result is that the maximum likelihood
only improves by $2\Delta \ln L = 5$ with one extra parameter and the
difference from $\Lambda$CDM is $2\Delta \ln L=23$, which is still
$4.8\sigma$ discrepant with most of the difference arising from the
changes to growth.

Furthermore, while the excess power at large angles can be reduced by
changing the initial power spectrum to eliminate large scale power,
existing WMAP5 CMB polarization measurements already forbid this
possibility.  Specifically, by introducing any finite cut off to the
initial power spectrum to flatten the temperature power spectrum, the
global likelihood decreases.

While it is still possible that the resolution of the ghost and strong
coupling issues of the theory can alter these consequences, it is
difficult to see how they can do so without altering the very mechanism
that makes it a candidate for acceleration without dark energy.  The
failure of this model highlights the power of combining growth and
distance measures in cosmology as a test of gravity on the largest
scales.

\begin{acknowledgments}

We thank K. Koyama and Y.S. Song for useful conversations. This work
was supported in part by the NSF grant AST-05-07161, by the
Initiatives in Science and Engineering (ISE) program at Columbia
University, and by the Pol\'anyi Program of the Hungarian National
Office for Research and Technology (NKTH).  SW and WH were supported
by the KICP under NSF PHY-0114422 WH was further supported by
U.S.~Dept. of Energy contract DE-FG02-90ER-40560 and the David and
Lucile Packard Foundation. LH was supported by the DOE grant
DE-FG02-92-ER40699, and thanks Alberto Nicolis for useful
discussions and Tai Kai Ng at the Hong Kong University of Science
and Technology for hospitality. MM was supported by the DOE grant
DE-AC02-98CH10886. Computational resources were provided by the
KICP-Fermilab computer cluster.

\end{acknowledgments}

\vfill
\bibliographystyle{physrev}
\bibliography{my}

\onecolumngrid
\appendix
\begin{center}
  {\bf APPENDIX A}
\end{center}

In Appendix A, we give details of the modification to CAMB employed
in the main text to calculate the modified growth of perturbations
in the DGP model under the PPF prescription.  For perturbations of
both the metric and matter in the various gauges, we use the same
notations as in Rfn.~\cite{Hu08}.

\subsection{PPF Description of Modified Gravity in the Linear Regime}

Structure formation in the linear regime for a certain class of modified
gravity theories can be equivalently described as an effective dark
energy component under ordinary gravity.  The requirements of this class
are that members remain metric theories in a statistically homogeneous
and isotropic universe where energy-momentum is conserved.  This class
includes the self-accelerating branch of the DGP model.

Following Rfns.~\cite{HuS07, Hu08}, scalar perturbations for the DGP
self-accelerating scenario can be parameterized by three free
functions, $g(\eta,k)$, $f_{\zeta}(\eta)$, and $f_G(\eta)$, and one
free parameter $c_{\Gamma}$.  We shall see in \S \ref{subsec:ppfDE}
that this is equivalent to specifying relationships between the
density perturbation, velocity and anisotropic stress of the
effective dark energy that close the conservation laws of the
effective dark energy.

The metric ratio or anisotropic stress parameter $g(\eta,k)$ is
defined as
\begin{equation}
\Phi_+ \equiv g(\eta,k) \Phi_- - {4\pi G \over H^2 k_H^2}
P_T\Pi_{T}\,,
\end{equation}
where $\Phi_+ \equiv (\Phi+\Psi)/2$, $\Phi_- \equiv (\Phi - \Psi)/2$,
with $\Phi=\delta g_{ij}/2g_{ij}$, $\Psi=\delta g_{00}/2g_{00}$ as the space-space and time-time metric
perturbations in the Newtonian gauge,
 $k_H=(k/aH)$, $\Pi$ stands for the
anisotropic stress, and the subscript ``$T$'' denotes sum over all the
true components. When anisotropic stresses of the true components are
negligible, $g$ just parameterizes deviations from GR in the metric
ratio of $\Phi_+$ to $\Phi_-$.

There are two key features of the PPF parameterization and these
define the two additional functions $f_{\zeta}(\eta)$ and
$f_G(\eta)$. The first is that the curvature perturbation in the
total matter comoving gauge $\zeta$ is conserved up to order $k_H^2$
in the super-horizon (SH) regime in the absence of non-adiabatic
fluctuations and background curvature
\begin{equation}
\lim_{k_H \ll 1} {\dot{\zeta} \over aH} = -{\Delta P_T - {2\over 3}
c_K P_T \Pi_T \over \rho_T + P_T} - {K \over k^2} k_H V_T + {1 \over
3} c_K f_\zeta(\eta) k_H V_T \,, \label{eqn:sh}
\end{equation}
where $f_{\zeta}(\eta)$ parameterizes the relationship between the
metric and the matter.  The second is that in the quasi-static (QS)
regime, one recovers a modified Poisson equation with a potentially
time dependent effective Newton constant
\begin{equation}
\lim_{k_{H}\gg 1} \Phi_- = {4\pi G \over c_K k_H^2 H^2}{\Delta_T
\rho_T + c_K P_T \Pi_T \over  1+f_G(\eta)} \,. \label{eqn:qs}
\end{equation}
Note that even if $f_G=0$, the Poisson equation for $\Psi$ may also
be modified by a non-zero $g(\eta,k)$. In the above two equations,
$\Delta$, $V$ and $\Delta P\,(\neq P\Delta)$ are density, velocity,
pressure perturbations in the total matter comoving gauge, and
$c_K=1-3K/k^2$. To bridge the two regimes, an intermediate quantity
$\Gamma$ is introduced,
\begin{equation}
\Phi_-+\Gamma = {4\pi G \over c_K k_H^2 H^2} (\Delta_T \rho_T + c_K
P_T \Pi_T)\,,
\label{eqn:phim}
\end{equation}
and an interpolating equation is adopted to make sure it dynamically
recovers the behavior specified by Eqs.~(\ref{eqn:sh}) and
(\ref{eqn:qs})
\begin{equation}
(1 + c_\Gamma^2 k_H^2) \left[{\dot{\Gamma} \over aH} + \Gamma +
c_\Gamma^2 k_H^2 (\Gamma - f_G \Phi_-)\right] =
S\,.
\label{eqn:dgamma}
\end{equation}
Here the free parameter $c_{\Gamma}$ gives the transition scale between
the two regimes in terms of the Hubble scale, and the source term $S$ is
given by,
\begin{equation}
S={\dot{g}/(aH) -2 g \over g+1} \Phi_- +  {4\pi G \over (g+1)k_H^2
H^2} \left\{ g \left[{\dot{(P_T\Pi_T)}\over aH} + P_T\Pi_T\right] -
\left[ (g + f_\zeta + g f_\zeta)(\rho_T+P_T) - (\rho_e+P_e) \right]
k_H V_T \right\}\,.
\label{eqn:source}
\end{equation}
By interpolating between two exact behaviors specified by functions of
time alone $f_\zeta(\eta)$ and $f_G(\eta)$, the PPF parameterization 
is  both simple and general.  Moreover, the one function of time and scale
$g(\eta,k)$ also interpolates between two well defined functions of time alone
for many models, including DGP.  The same is not true of parameterizations
that involve the effective anisotropic stress alone or the metric functions directly.

\subsection{PPF Parameterization for DGP}

The PPF parameterization for the self-accelerating DGP is given in
Rfn.~\cite{HuS07}, which we summarize as the follows.  On super-horizon
scales, the iterative scaling solution developed in Rfn.~\cite{SaSH07}
is well described by
\begin{equation}
g_{\rm SH}(\eta)={ 9 \over 8 H r_c -1} \left( 1 + {0.51 \over H r_c
- 1.08} \right) \,, \label{eqn:gsh}
\end{equation}
and $f_\zeta(\eta)=0.4 g_{\rm SH}(\eta)$, while in the QS regime,
the solution is parameterized by \cite{KoM06}
\begin{equation}
g_{\rm QS}(\eta) = -{1\over 3}\left[ 1-2 H r_c\left( 1 + {\dot{H}
\over 3aH^2}\right) \right]^{-1}\,,
\end{equation}
and $f_G(\eta)= 0$.  On an arbitrary scale in the linear regime, $g$
is then interpolated by
\begin{equation}
g(\eta,k) = { g_{\rm SH} +  g_{\rm QS}(c_{g}k_H)^{n_{g}} \over 1+
(c_{g}k_H)^{n_{g}}}\,,
\end{equation}
with $c_g=0.14$ and $n_g=3$. This fitting formula has been shown to
give an accurate prediction for the evolution of $\Phi_-$ according
to the dynamical scaling solution \cite{HuS07}. The transition scale
for $\Gamma$ is set to be $c_\Gamma=1$, {\it i.e.}~at the horizon
scale.  We note here that the solutions developed in
Rfns.~\cite{KoM06,SaSH07} are for flat universes, so strictly
speaking, the above PPF parameterization only works for DGP with
$\Omega_K \rightarrow 0$.  However for the small curvatures that are
allowed by the data, we would expect its effect on structure
formation to be small and arise from terms such as $H^2 \rightarrow
H^2+K/a^2$ (see, {\it e.g.}, Rfn.~\cite{GiSK08}).  Given the cosmic
variance of the low-$\ell$ multipoles, these corrections should have
negligible impact on the results.

\subsection{``Dark Energy'' Representation of PPF}
\label{subsec:ppfDE}

By comparing the equations that the PPF quantities satisfy with their
counterparts in a dark energy system under general relativity, we obtain
the following relations for the perturbations of PPF's corresponding
effective dark energy.  These relations act as the closure conditions
for the stress energy conservation equations of the effective dark
energy.  The first closure condition is a relationship between the PPF
$\Gamma$ variable and the components of the stress energy tensor of the
effective dark energy
\begin{equation}
\rho_{e}\Delta_{e} + 3(\rho_{e}+P_e) {V_{e}-V_{T}\over k_{H} } +
c_K P_e\Pi_{e} = -{k^2 c_K \over 4\pi G a^2} \Gamma \,.
\end{equation}
The second closure condition is a relationship for the anisotropic
stress
\begin{equation}
P_e\Pi_e=-\frac{k_H^2H^2}{4 \pi G}g\Phi_-\,.
\label{eqn:pie}
\end{equation}
Stress energy conservation then defines the velocity perturbation
\begin{equation}
{ V_{e}-V_{T}\over k_{H}} =-{H^2 \over 4\pi G (\rho_{e} + P_e) }
{g+1 \over F} \left[ S - \Gamma - {\dot{\Gamma} \over aH} +
f_{\zeta}{4\pi G (\rho_{T}+P_T) \over H^2}{V_{T}\over k_{H}}
\right] \,,
\label{eqn:Ve}
\end{equation}
with
\begin{equation}
F = 1 + {12 \pi G a^2 \over k^2 c_K} (g+1)(\rho_T + P_T)\,.
\end{equation}
For details of all the derivations, see Rfn.~\cite{Hu08}.  The PPF
equation for $\Gamma$ then replaces the continuity and Navier-Stokes
equations for the dark energy.  Note that the effective dark energy
pressure perturbation, which obeys complicated dynamics to
enforce the large and small scale behavior, is not used.

\subsection{Modifying CAMB to Include PPF}

Given the dark energy representation of PPF, we only need to modify
the parts in CAMB where dark energy perturbations appear explicitly
or are needed, in addition to its equation of state which will be
specified by the desired background expansion. These include the
Einstein equations and the source term for the CMB temperature
anisotropy. Since CAMB adopts the synchronous gauge, in this
section, we will express everything in the same gauge.

The two Einstein equations used in CAMB are,
\begin{eqnarray}
{\dot{h_L}\over 2k} - c_K k_H \eta_T &=& {4\pi G \over k_H H^2}
(\rho_{T}\delta_{T}^{s} + \rho_{e} \delta_{e}^{s} )\,,\\
k\dot{\eta_{T}}- K {( \dot{h_L} + 6\dot{\eta_{T}}) \over 2k} &=&
{4\pi G a^2} \left[ (\rho_{T}+P_T)v_{T}^s +
(\rho_{e}+P_e)v_{e}^s \right] \,,
\end{eqnarray}
where $h_L$, $\eta_T$ are metric perturbations, $\delta$, $v$ are
density and velocity perturbations, and superscript ``s'' labels the
synchronous gauge. Here, we only need to provide $\delta_e^s$ and
$v_e^s$.  Given the gauge transformation relation for velocity
\begin{equation}
V=v^s+k\alpha\,,
\end{equation}
with $\alpha \equiv (\dot{h_L} + 6\dot{\eta_{T}}) / 2k^2$, the
following expression for $v_e^s$ is easily obtained from
Eq.~(\ref{eqn:Ve})
\begin{equation}
(\rho_e+P_e)v_e^s = (\rho_e+P_e)v_T^s - \frac{k_H H^2}{4 \pi G}
\frac{(g+1)}{F}  \left[ S - \Gamma - \frac{\dot{\Gamma}}{aH} +
f_{\zeta}\frac{4\pi G(\rho_T+P_T)(v_T^s + k \alpha )}{k_H H^2}
\right]\,.
\label{eqn:ve}
\end{equation}
In order to calculate $v_e^s$, we need to evaluate $\alpha$, which we
find, with the help of the first closure condition and the two Einstein
equations, to be given by
\begin{equation}
k \alpha = k_H \eta_T + \frac{4 \pi G}{c_K k_H H^2} \left[\rho_T
\delta_T^s + \frac{3(\rho_T+P_T)v_T^s}{k_H}\right] - \frac{4 \pi G
}{k_H H^2}P_e \Pi_e - k_H \Gamma \,.
\end{equation}
Here, the anisotropic stress is gauge-independent, and is given by
Eq.~(\ref{eqn:pie}), where $\Phi_-$ is given by gauge-transforming
the density perturbation in Eq.~(\ref{eqn:phim}) according to
\begin{equation}
\rho \Delta = \rho \delta^s - \dot{\rho} {v_T^s \over k}\,.
\end{equation}
In addition to $\alpha$, we need also to specify $S$ and $\dot{\Gamma}$
in order to get $v_e^s$. Given $\Phi_-$, $S$ can be calculated by
gauge-transforming $V_T$ in Eq.~(\ref{eqn:source}), and $\dot{\Gamma}$
then follows from Eq.~(\ref{eqn:dgamma}).  Provided $v_e^s$ and
$P_e\Pi_e$, $\delta_e^s$ can be obtained by gauge-transforming the first
closure condition
\begin{equation}
\rho_e\delta_e^s=- c_K P_e \Pi_e - 3(\rho_e + P_e) {v_e^s \over k_H}
-\frac{c_K k_H^2 H^2}{4 \pi G} \Gamma\,.
\end{equation}

The source term for the CMB temperature anisotropy is given by
\cite{ZaSB97}
\begin{equation}
S_T(\eta, k)={\cal G}\left(\Delta_{T0}+2 \dot{\alpha} +{\dot{v_b^s}
\over k}+{\Sigma \over 4 {\bar b}} +{3\ddot{\Sigma}\over 4k^2 {\bar
b}}\right) + e^{-\kappa}(\dot{\eta_T}+\ddot{\alpha}) +\dot{\cal
G}\left({v_b^s \over k}+\alpha +{3\dot{\Sigma}\over 2k^2 {\bar
b}}\right) +{3 \ddot{\cal G}\Sigma \over 4k^2 {\bar b}}\,,
\end{equation}
where the visibility function ${\cal G} \equiv -\dot{\kappa}
\exp(-\kappa)$, with $\kappa(\eta)$ the optical depth from today to
$\eta$ for Thomson scattering, $\Delta_{T \ell}$ is the multipole of the
temperature anisotropy, $v_b^s$ is the baryon velocity, $\Sigma \equiv
\Delta_{T2} -12 _2\Delta_{P2}$, with $_2\Delta_{P2}$ the quadrupole of
the polarization anisotropy (for more explicit definitions for
$\Delta_{T \ell}$ and $_2\Delta_{P \ell}$, see Rfn.~\cite{ZaSB97}), and
$\bar{b}^2=c_K$.  Here we need to specify $\dot{\alpha}$ and
$\ddot{\alpha}$. $\dot{\alpha}$ is given by the Einstein equation
\begin{equation}
\dot{\alpha}+2{\dot{a} \over a} \alpha - \eta_T = -{8 \pi G \over
k_H^2 H^2} (P_T \Pi_T + P_e \Pi_e) \,,
\end{equation}
Differentiating this equation with respect to $\eta$ also gives us
$\ddot{\alpha}$, which will need the derivative of the effective dark
energy's anisotropic stress.  From energy-momentum conservation and
the first closure condition, we obtain
\begin{equation}
\dot{(P_e \Pi_e)}={\dot{h_L} \over 2} {{(\rho_e + P_e)} \over c_K} -
\frac{k_H^2 H^2}{4 \pi G} \dot{\Gamma} + {aH \over c_K}  \left\{
\delta \rho_e^s + (\rho_e+P_e) v_e^s \left[ {6 \over k_H} + k_H - {3
\over k}\left({\dot{a} \over a }+{\dot{H} \over H }\right)\right]
\right\}\,,
\end{equation}
which completes our modification of CAMB.

\onecolumngrid
\appendix
\begin{center}
  {\bf APPENDIX B}
\end{center}

\begin{figure}[t!]
\resizebox{90mm}{!}{\includegraphics{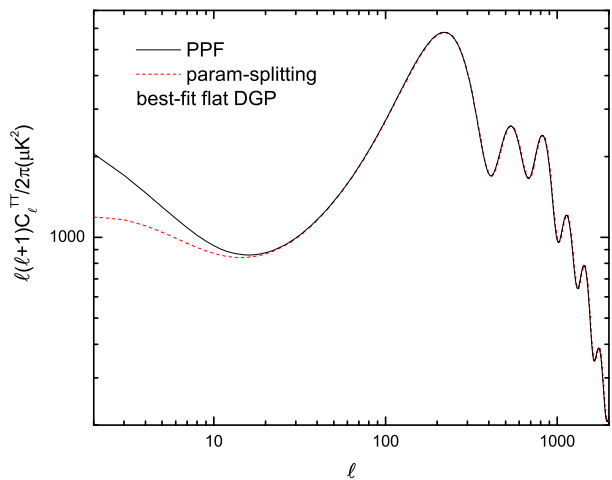}}
\caption{\label{fig:paramsplit} A comparison of the PPF prediction
(upper solid line) with the prediction by parameter-splitting (lower
dashed line) for the DGP model. }
\end{figure}

In Appendix B, we briefly contrast the PPF prediction for the CMB
temperature power spectrum with an attempt to approximate the DGP
prediction via the parameter-splitting technique
\cite{ZHS05,WaHMH07}. The technique works by splitting the dark
energy equation of state $w$ into two separate parameters, with one,
$w_{\rm geometry}$, determining geometric distances and the other,
$w_{\rm growth}$, determining the growth of structure. We choose
$w_{\rm geometry} = -0.7$ and $w_{\rm growth} = -0.57$, which are
obtained respectively by fitting to $H(z)$ (for $z = 0$ to $10$) and
fitting to the quasi-static growth factor (for $z = 0$ to $2$)
according to the best-fit flat DGP model (Table \ref{tab:fbf}). The
result is shown in Fig. \ref{fig:paramsplit}. One can see that
parameter-splitting falls short of the PPF prediction on large
angular scales. This illustrates the importance of correctly
modeling the perturbation growth on horizon scales. The
parameter-split of $w$ does not contain enough freedom to describe
perturbation growth in both the sub-horizon (quasi-static) and
super-horizon regimes.

\end{document}